\documentstyle[12pt,epsfig]{article}
\begin{document}
\begin{titlepage}
\begin{center}
\vspace*{3cm}

%\begin{title}
{
\Large  \bf Are charge fluctuations a good signal for 
QGP?
}
%\end{title}
\vspace{2cm}

\begin{author}
\Large
K. Fia{\l}kowski\footnote{e-mail address: uffialko@thrisc.if.uj.edu.pl},
R. Wit\footnote{e-mail address: wit@thrisc.if.uj.edu.pl}

\end{author}

\vspace{1cm}

{\sl M. Smoluchowski Institute of Physics\\ Jagellonian University \\

30-059 Krak{\'o}w, ul.Reymonta 4, Poland}

\vspace{3cm}

\begin{abstract}
A recent proposal to study charge fluctuations as a  possible signal of 
quark - gluon plasma is discussed. It is shown 
that the "pion gas model" considered as the reference sample is unrealistic 
and the expected signal from plasma may be difficult to distinguish from 
that coming from "standard collisions".
\end{abstract}

\end{center}
\vspace{1cm}

%PACS:  \\

%{\sl Keywords:}  Fluctuations, hadronic gas, quark gluon plasma  \\

\vspace{1cm}

\noindent

 2 June, 2000 \\

\end{titlepage}

\par
In a recent paper Jeon and Koch \cite{JK} argued that the event-by-event 
fluctuations of charge (or, equivalently, of the ratio of positive to 
negative pions) in a restricted rapidity range may be used as a tool to signal 
the 
possible formation of quark - gluon plasma (QGP). They compare the ratio of 
charge 
dispersion squared to the charged multiplicity in two models: the "pion gas" and 
QGP 
and conclude that they differ by a factor of five. Similar result is found for 
the 
dispersion of "positive-to-negative" ratio for pions. The dramatic difference 
may be 
easily  understood as the reflection of small quark charges as compared to 
hadrons 
and of zero gluon charges.
\par
The Authors conclude that the strong decrease of charge fluctuations in the QGP 
as 
compared to a pion gas should be seen as an "unmistakable signal of QGP 
formation from 
'Day-1' measurements" at RHIC. They add later some caveat about resonances and 
other 
correlation effects which may reduce the fluctuations in the "pion phase", but 
these 
are claimed to be minor corrections.
\par
However, the value of ratio $D_Q^2/<n_{ch}>$ obtained in the pion gas model 
seems to be unrealistic. In the past charge fluctuations were measured in hadron
collisions, usually for separated CM hemispheres \cite{ABB}. The values of 
$D_Q^2$
were systematically much lower than $<n_{ch}>$; semi-inclusive data
indicated that charge dispersion squared is indeed proportional to charged
multiplicity, but the proportionality coefficient is below $0.5$.
\par
For the energy and rapidity range similar to that considered in \cite{JK} (and 
expected at RHIC) a reliable estimate is available from the JETSET/PYTHIA 
generator
\cite{SJO} for $pp$ collisions. We performed such calculations at CM energy of
 180 GeV and 1800 GeV
for centrally located bins of rapidity of width 2, 4, 6 and 8. The results for 
the ratio
$D_Q^2/<n_{ch}>$ as a function of a bin width are shown in Fig.1. 

\begin{figure}[h]
\centerline{
\epsfxsize=11cm
\epsfbox{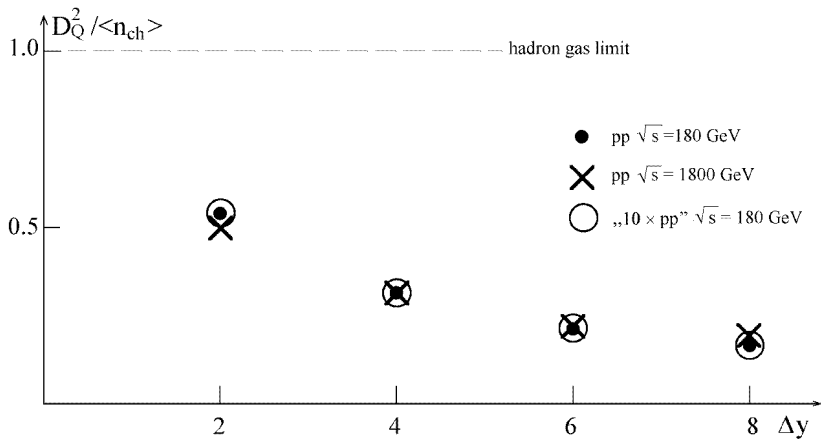}
}
\caption{\label{fig1} {\sl The values of the ratio $D_Q^2/<n_{ch}>$ from
the PYTHIA/JETSET generator for $pp$ collisions at 180 GeV (dots) 
 and  the  1800 GeV (crosses) vs half bin width in rapidity. The values 
 for "heavy ion events" formed from groups of ten $pp$ events at 180 GeV are 
shown as 
 open circles.} } 
%\vspace{0.4cm}
\end{figure}

One sees a fast decrease
with the bin width; the values are changing between $0.5$ and $0.15$. This 
reflects the approximate 
 ordering of charges (with alternating signs) along a fragmenting string in the 
model; charge dispersion 
squared increases much slower than average multiplicity, and for wide bins it is 
bound 
in fact to decrease towards a limiting value of zero (when all the phase-space 
is covered).
The energy dependence is very weak. This is also not surprising:  for the 
central 
region in the PYTHIA generator increasing energy results just in the growing 
number of contributing strings. This 
increases in very similar way both the numerator and denominator of the 
ratio.

\par
Obviously, we need an estimate for heavy ion collisions and not for hadron 
scattering.
However, since our aim is to contrast the QGP signal with "standard" collisions,
it should be instructive to superimpose many $pp$ collisions to mimick a heavy 
ion
interaction. We have done it using the momenta from groups of ten $pp$ events to 
form 
single "events". The ratio $D_Q^2/<n_{ch}>$ resulting from this simulation is 
also
shown in Fig.1. As expected, the points are indistinguishable from those for 
$pp$
collisions: both numerator and denominator of the ratio are simply multiplied by
the number of superimposed events, leaving the value of the ratio unchanged.
\par
We see that the value of the investigated ratio for hadron collisions in a 
realistic 
model without QGP is much smaller than $1$ and similar results are expected for 
heavy 
ion collisions, if describing them as superposition of independent 
nucleon-nucleon
collisions is a reasonable approximation. Moreover, the value depends quite 
strongly
on the chosen range of rapidity. It may be easily as low as the prediction from 
QGP.
Thus a measurement of this ratio at RHIC is not likely to prove or disprove the 
formation of QGP. Obviously, for more precise calculations of the "reference 
values 
without QGP" one should use the generators dedicated for heavy ion collisions 
(as 
FRITIOF or VENUS). However, we do not see any reasons why they should give 
results different from ours.
\par
Summarizing, we have shown that the predictions for the ratio $D_Q^2/<n_{ch}>$
from the "pion gas model" used in [1] as reference for the possible signal of 
QGP
are not reliable. The realistic estimates are much lower, which seems to 
diminish the value of 
this quantity as a sensitive indicator of the QGP formation.\\

{\bf Acknowledgements}
\par
We would like to thank K. Zalewski for drawing our attention to ref. 
\cite{JK} and A. Bia{\l}as for suggesting this investigation. This work is 
partially supported by KBN grants \# 2 P03B 086 14 and \# 2P03B 010 15. One 
of us (RW) is grateful for a partial financial support by the KBN grant \# 2 
P03B 019 
17.

\end{document}